\documentstyle[multicol,prl,aps,epsf,psfig,epsfig]{revtex}
\begin{document}
\title
{
Probability distribution of persistent spins in a Ising chain
}
\author
{
Pratap Kumar Das and Parongama Sen
}
\address
{
$^1$Department of Physics, University of Calcutta,
    92 Acharya Prafulla Chandra Road, Kolkata 700009, India.
}
\maketitle
\begin{abstract}
 We study the probability distribution $Q(n,t)$  of  $n(t)$, the
fraction of spins unflipped till time $t$, in a Ising chain with ferromagnetic
interactions. The distribution shows a peak at $n=n_{max}$ and in general 
is non-Gaussian and asymmetric in nature. However for $n>n_{max}$ it shows a
Gaussian decay. A data collapse can be obtained when 
$Q(n,t)/L^{\alpha}$ versus $(n-n_{max})L^{\beta}$ is plotted 
with $\alpha \sim 0.45$
and $\beta \sim 0.6$. Interestingly, $n_{max}(t)$ shows a  different behaviour
compared to $ \langle n(t)\rangle =
P(t)$, the persistence probability which follows  the well-known 
behaviour  $P(t)\sim t^{-\theta}$. A quantitative estimate of the asymmetry
and non-Gaussian nature of $Q(n,t)$ is made by calculating its skewness and
kurtosis.
\end{abstract}   
  
PACS nos: 05.70.Ln,05.50.+q,64.60.Ht

  Preprint no: CU-physics-05/2004.
  
\begin{multicols}{2}
In recent years a lot of work has been devoted in studying 'persistence' in 
dynamical systems \cite{Satya_rev}.  
Persistence is the phenomenon 
 defined as the probability that a fluctuating non-equilibrium
field has not changed its sign upto  time $t$. 
This phenomenon has been observed in magnetic systems 
\cite{Derrida,Stauffer,Maj_Sire,Satya_Sire,Sen}, simple diffusion 
\cite{Maj_Bray}, coarsening dynamics \cite{Bray},
various models undergoing phase separation process \cite{Krap_Redner},
fluctuating interfaces \cite{Krug_Cornell}  etc.

In Ising system, 
 persistence is simply the probability that a spin has not changed its 
 sign up to time 
$t$ after the system is quenched to a low temperature from an 
initial high temperature.
    The fraction of the persistent spins $P(t)$ here 
    is given by 
\begin{equation}
P(t){\sim}t^{-\theta},
\end{equation}
where $\theta$ is a new exponent not related to any previously known static or 
dynamic exponent. In one-dimension, at $T$=$0$, $\theta$ is exactly known, 
$\theta$=$0.375$ \cite{Derrida}.  
In higher dimensions, the persistence exponent has been obtained
approximately using both analytical and numerical methods. 

    In the numerical
studies, one needs to generate different
random initial configurations to obtain the persistence probability
$P(t)$ which is  an averaged out quantity. We define $n(t)$ 
to be the fraction of persistent spins till time $t$ which has different values 
for different
realisations of randomness such that $\langle n(t) \rangle = P(t)$, where $\langle\rangle$ 
denotes average over realisations. Thus  
$n(t)$ can be  defined as a stochastic variable
described by a probability distribution function.
We have precisely studied the probability distribution $Q(n,t)$  of $n(t)$
and  obtained a number of interesting features of the distribution 
for the one-dimensional Ising model. Probability distribution of random variables
makes interesting studies in various systems, e.g., for the mass of spanning 
cluster in percolation \cite{Stau_Ah}, random Ising and bond diluted
Ashkin-Teller model \cite{Wise}, conductance of classical dilute resistor
network \cite{Lub}, directed polymers and growth models \cite{Kim},
degree distribution in networks \cite{New} etc. 
In several of these systems, the distribution is non-Gaussian
and shows many interesting features. Certain properties like 
self averaging, multifractality  etc.
can be studied directly from the distribution function \cite{Ah_Ha}.
Also measurements like skewness and kurtosis \cite{Kurtosis} from the higher moments to 
estimate quantitatively the asymmetry and departure from Gaussian behavior
of the distribution are possible.

We have considered a chain of Ising spins with nearest-neighbour
ferromagnetic interaction  and simulated it using periodic boundary 
conditions. The interaction is represented by the Hamiltonian
\begin{equation}
H=-J\sum_{i}s_{i}s_{i+1}.     
\end{equation}
The initial configuration is random and single spin flip (deterministic) Glauber dynamics has been
used for subsequent updating. 
    
\begin{center}
\begin{figure}
\noindent \includegraphics[clip,width= 5cm, angle=270]{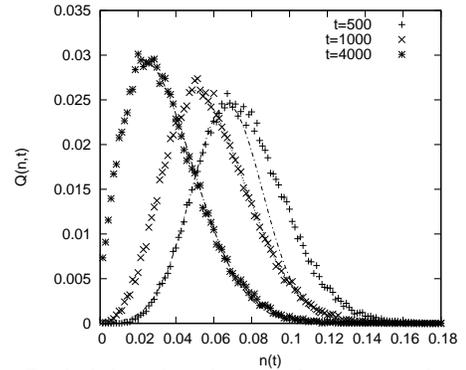}
\caption{Probability distribution $Q(n,t)$ as a function of $n$ at $t$ = 500,
1000 and 4000 for $L = 700$. The continuous lines are Gaussian  fits  
of the form  $\exp(-(x-x_0)^2/{\alpha})$. For the fits  shown  for $t$ = 
1000 and
4000 on the  right side of the peak,  $\alpha= \alpha_r= 1.5\times 10^{-3}$ and $1.4\times 10^{-3}$ respectively. 
The left side of the peak can be fitted to the same form  with a different
value of $\alpha$: 
$\alpha=\alpha_{l}=7.0\times 10^{-4}$ as shown in the curve for $t$ =500.}
\end{figure}
\end{center}
\begin{center}
\begin{figure}
\noindent \includegraphics[clip,width= 5cm, angle=270]{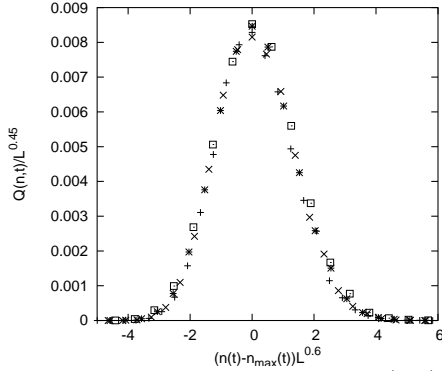}
\caption{Scaled probability distribution $Q(n,t)/L^{0.45}$ as a function of 
$(n(t)-n_{max})L^{0.6}$ at $t$ = 200
 for $L$ = 500,600,700 and 1000 shows a collapse.}
\end{figure}
\end{center}
Since a large number of configurations are required to get accurate data,
the  system sizes are  restricted to $L\leq 1000$. Primarily, we are 
interested in the form of the distribution $Q(n,t)$ which is plotted at
different times against $n(t)$ in Fig $1$. We have obtained $Q(n,t)$
for several values of $L$ also.  

       At all times, the distribution $Q(n,t)$ shows a peak at $n(t)=n_{max}(t)$. 
The value of $n_{max}(t)$ has negligible dependence on $L$. At early times, there is a
Gaussian decay of $Q(n,t)$ on both sides of the peak.
Interestingly, the Gaussian behavior 
$\exp(-(x-x_0)^2/{\alpha})$ is followed with different $\alpha$ values on the 
two sides of $n_{max}$, $\alpha_{l}$ on the left and  $\alpha_{r}$ on the right. 
For very large $t$, $\alpha_{r}$
is the only measure possible as it is difficult to 
fit the function to a Gaussian on the left side of the peak. 
Usually the decay behaviour of distributions for rare events is of interest and we find
the decay is Gaussian at all times for $n>n_{max}$. 
We observe that $\alpha_{r}$ shows a 
weak dependence on $t$ which  becomes negligible  for larger system 
sizes. It is also a function of $L$, $\alpha_{r}\sim L^{-1.2}$. 
$\alpha_{l}$, which can be calculated accurately for initial times, follows
a similar scaling.
In fact, the scaled distribution $Q(n,t)/L^{\alpha}$ plotted against 
$(n(t)-n_{max}(t))L^{\beta}$ with $\alpha \sim 0.45$ and $\beta \sim 0.6$ shows a nice 
data collapse (Fig 2). Even at long times, a fairly good data 
collapse can be obtained with these values of $\alpha$ and $\beta$.

           The distribution has natural cut-offs at $n=0$ 
and $n=1$. Therefore as time evolves, the distribution becomes more and more 
asymmetric as the fraction of persistent spins decreases with time. This 
asymmetry is more apparent when the probability that 
there is no persistent spin, 
 $Q(0,t)$,  begins to assume finite values.

 In fact the behaviour of  $Q(0,t)$ is quite interesting itself.
 In Fig 3, $Q(0,t)$ has been plotted against $t/L^2$ and
 the data for different $L$ values seem to fall on the same curve indicating
 that  $Q(0,t)$ is a function of
  $t/L^{2}$ with the  behavior  
\[
  Q(0,t)=0   ~~{\rm for}~~  t/L^{2}<a_{0}
\]
\[
  ~~~~~~~~\neq 0 ~~{\rm for}~~  t/L^{2}>a_{0}
\]
      where $a_{0}\sim 0.001$.
      While $t/L^2$ appearing as a scaling argument is expected,
      what is noticeable is the small value of the threshold $a_0$.

\begin{center}
\begin{figure}
\noindent \includegraphics[clip,width= 5cm, angle=270]{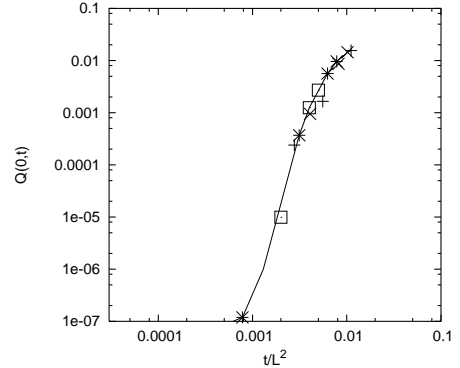}
\caption{$Q(0,t)$ as a function of $t/L^2$ for four different $L$ values.
The solid line is a guide to the eye.}
\end{figure}
\end{center}

       The comparison of the behaviour of the most probable value $n_{max}(t)$
and the average value $P(t)$ shows consistent features.
In Fig $4$, we have plotted both $P(t)$ and $n_{max}(t)$ against $t$. While 
$P(t)$ shows the expected power law decay with the known exponent 
$\theta=0.375$, $n_{max}(t)$ shows a different behavior. $n_{max}(t)$ falls off
faster than $P(t)$ and it is not possible to 
fit a power law  to it.
\begin{center}
\begin{figure}
\noindent \includegraphics[clip,width= 5cm, angle=270]{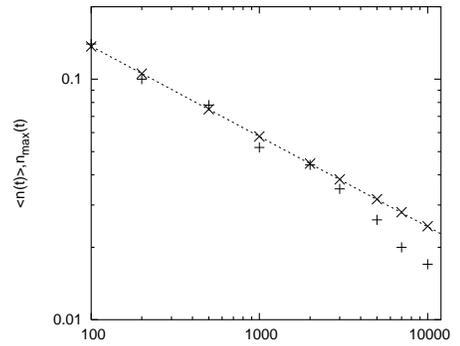}
\caption{ $\langle n(t)\rangle=P(t)$ ($\times$) and $n_{max}(t)$ ($+$) are plotted against time $t$ 
for $L=1000$. While $\langle n(t)\rangle(t)\sim t^{-0.375}$ 
$n_{max}(t)$ shows deviation from this behavior at finite times (the dashed 
line 
with slope -$0.375$ is fitted to $\langle n(t)\rangle$).} 
\end{figure}
\end{center}
 One can  try a power law fit only for very early times
with a value of $\theta$ close to 0.375.
For $P(t)$  it is known that the behavior 
$P(t)\sim t^{-\theta}$ is valid for $t<\tau$, where $\tau\sim L^2$.
In case the power law behaviour of  $n_{max}$ is valid for a finite time 
in the same sense, it appears that the deviation from the power law takes place
at a much  earlier time.
We find that the so called deviation occurs at a value of $t/L^2 \sim a_0$
indicating that the most probable value fails to show persistence
behaviour when $Q(0,t)$ becomes non-zero. Also $P(t)$ for $t \to \infty $ goes to 
constant value ($\sim L^{-2\theta}$) while $n_{max}$ goes to zero for $t\to \infty$.

\begin{center}
\begin{figure}
\noindent \includegraphics[clip,width= 5cm, angle=270]{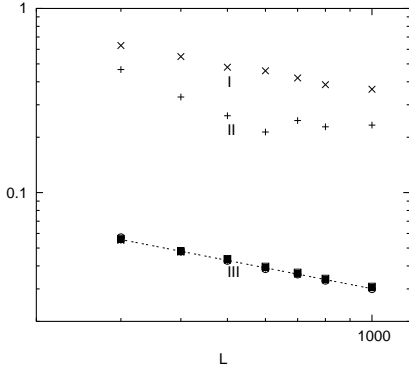}
\caption{The skewness (curve marked I), kurtosis (II) and scaled  
 $R_n$ (III)
 at $t=500$ as a function of the system size $L$ have been shown.
  $R_{n}/t^{\theta_t}$ (data for 4 different values of time
have been plotted) is seen to vary as
 $L^{-\theta_l}$ with $\theta_l \sim 0.51$ and $\theta_t \sim 0.345$. }
\end{figure}
\end{center}

One can easily calculate the higher moments from  the probability
distribution.
We have studied the self averaging property and also tried to estimate the 
asymmetry and the non-Gaussian behavior by calculating the skewness and kurtosis 
of the distribution.

         A system is said to exhibit self averaging if 
$R_{x}(L)=(\Delta x)^2/\langle x\rangle ^{2} \to 0$ as $L \to \infty$ 
for any physical quantity $x$ \cite{Wise,Ah_Ha} ($\Delta x $ is the 
variance $=\langle x^2\rangle-\langle x\rangle ^{2}$.) Here we have calculated 
$R_{n}(L)=(\Delta n) ^{2}/
\langle n(t)\rangle ^{2}$ to check whether self-averaging is present. 
Our results show that $R_{n}(L,t) \sim L^{-\theta_{l}}$ 
where $\theta_{l} \sim 0.51$ indicating strong self averaging. In fact 
$R_{n}(L,t)$ also shows a power law increase with time such that 
$R_{n}(L,t)/t^{\theta_{t}}$ for different values 
of $L$ shows a collapse (Fig $5$). Apparently, the variance is a weak function 
of time: $(\Delta n)^{2}\sim t^{-0.06}$.

     To measure the skewness, we calculate
       $s(t,L)=M_{3}/(M_{2})^{3/2}$  
  where the $m$th centered moment is $M_{m}=\langle(x-\langle x\rangle)^{m}\rangle$.
$s$ measures the asymmetry of the distribution and is zero for a 
symmetric distribution. Here it shows an increase with time
as expected (e.g for $L=1000$
 $s\sim $0.18, 0.36 and 0.57 for $t=$ 200, 500 and 2000 respectively),  but 
shows a decay with $L$ as well. However, the dependence on $L$
weakens at longer times. We expect $s$ to remain finite at $L\to\infty$, which
is not so apparent from the data presumably because of the small system 
sizes considered.  
That there is no universal distribution with respect to time
as in $[10]$ is reflected by fact that $s$ is time dependent.

      Kurtosis is a measure of the peakedness of the distribution.
 It is studied by calculating $k(t,L)=M_{4}/(M_{2})^{2} - 3$.
For a Gaussian distribution, $k(t,L)=0$ indicating that the peak is at the mean 
value. A negative value of $k$ would imply that the distribution is flat.
$ k$ shows a saturation with $L$ for all times indicating the non-Gaussian 
behavior of the distribution. It has a positive value $\sim 0.2 $ 
to show that the distribution is peaked close to the mean. This value is also 
independent of time implying that non-Gaussian behavior remains constant 
quantitatively with time. In Fig 5, typical variation of $s$ and $k$ 
with $L$ have been shown. 
    
   In summary, we have obtained the distribution function for the fraction of 
persistent spins $n(t)$ for a one
dimensional ferromagnetic Ising system. The form of the distribution 
is non-Gaussian in general. The form also changes with time, 
becoming more and more asymmetric at longer times.
The most probable value $n_{max}(t)$ shows deviation from the 
average $\langle n(t)\rangle = P(t) \sim t^{-\theta}$ at times $t/L^{2}>a_{0}$
 where $a_{0} \sim 10^{-3}$. Here we also find that $Q(0,t)$ begins 
to take non-zero values. The system also shows strong self averaging.

Acknowledgments: P.K. Das acknowledges support from CSIR grants no.
9/28(608)/2003-EMR-I. P. Sen acknowledges support from DST grant no. 
SP/S2/M-11/99.
\vskip 1cm
Email:  {pra$\_$tapdas}@rediffmail.com, parongama@vsnl.net

\end{multicols}
 

\begin{thebibliography}{references:}

\bibitem{Satya_rev} For a review, see S.N. Majumdar, {Curr. Sci. {\bf{77}} 370 (1999)} 
 and the references therein.
\bibitem{Derrida} B. Derrida, A.J.Bray and C. Godreche, {J.Phys. A {\bf{27}} L357 (1994)};
 B. Derrida, V. Hakim and V. Pasquier, {Phys. Rev. Lett. {\bf{75}} 751 (1995)}.
\bibitem{Stauffer} D. Stauffer, {J. Phys.A {\bf{27}} 5029 (1994)}.
\bibitem{Maj_Sire} S.N. Majumdar and C. Sire, {Phys. Rev. Lett. {\bf{77}} 1420
(1996)}.
\bibitem{Satya_Sire} S.N. Majumdar, A.J. Bray, S.J. Cornell and C. Sire, {Phys. 
Rev. Lett. {\bf{77}} 3704 (1996)}.
\bibitem{Sen} M. Saharay and P. Sen, {Physica A {\bf{318}} 243 (2003)}.
\bibitem{Maj_Bray} S.N. Majumdar and A.J Bray,  {Phys. Rev. Lett. {\bf{81}} 2626
 (1998)}.
\bibitem{Bray} A.J. Bray, B. Derrida and C. Godreche, {Europhys. Lett. {\bf{27}}
 175 (1994)}. 
\bibitem{Krap_Redner} P.L. Krapivsky, E.Ben-Naim and Redner, 
{Phys. Rev.E {\bf{50}} 2474 (1994)}.
\bibitem{Krug_Cornell} J.Krug, H.Kallabis, S.N.Majumdar, S.J.Cornell, A.J.Bray 
and C.Sire, {Phys. Rev.E {\bf{56}} 2702 (1997)}.
\bibitem{Stau_Ah} D. Stauffer and A. Aharony, {\it{Introduction to Percolation
Theory}} (Taylor and Francis, London, 1994); P. Sen, {Int. J. Mod. Phys.C {\bf{10}} 747 (1999)}.
\bibitem{Wise} S. Wiseman and E. Domany, {Phys. Rev.E {\bf{58}} 2938 (1998)}.
\bibitem{Lub} A.B. Harris and T.C. Lubensky, {Phy. Rev.B {\bf{35}} 6964 (1987)}.
\bibitem{Kim} J.M. Kim, M.A. Moore and A.J. Bray, {Phy .Rev.A {\bf{44}} 
2345 (1991)}.
\bibitem{New} R. Albert and A.L. Barab\'asi, {Rev. Mod. Phys. {\bf{74}} 47 (2002)}.
\bibitem{Ah_Ha} A. Aharony and A.B. Harris, {Phy. Rev. Lett. {\bf{77}} 3700 (1996)}.
\bibitem{Kurtosis} A. Abramowitz and I. Stegun, {\it{Handbook of Mathematical 
Functions}}, p.928 (Dover publications, New York, 1972).
 
\end{thebibliography}
\end{document}